\documentclass[a4paper,11pt]{article}
\usepackage{pos}

\title{What is chiral susceptibility probing?}

\author[a,b]{JLQCD Collaboration: S.~Aoki}
\author[c]{Y.~Aoki}
\author*[d]{H.~Fukaya}
\author[e,f]{S.~Hashimoto}
\author[d]{C.~Rohrhofer}
\author[g]{K.~Suzuki}

\affiliation[a]{
  Center for Gravitational Physics, Yukawa Institute for Theoretical Physics,
        Kyoto University, Kyoto 606-8502, Japan}

\affiliation[b]{RIKEN Nishina Center (RNC), Saitama 351-0198, Japan}

\affiliation[c]{RIKEN Center for Computational Science, 7-1-26
Minatojima-minami-machi, Chuo-ku, Kobe, Hyogo 650-0047, Japan}

\affiliation[d]{Department of Physics, Osaka University, 
        Toyonaka, Osaka 560-0043 Japan}

\affiliation[e]{High Energy Accelerator Research Organization (KEK),
        Tsukuba 305-0801, Japan}
\affiliation[f]{School of High Energy Accelerator Science,
        The Graduate University for Advanced Studies
        (Sokendai),Tsukuba 305-0801, Japan}
\affiliation[g]{Advanced Science Research Center, Japan Atomic
  Energy Agency (JAEA), Tokai 319-1195, Japan
  }

\abstract{In the early days of QCD, the axial $U(1)$ anomaly was considered as
a trigger for the breaking of the $SU(2)_L\times SU(2)_R$ symmetry through
topological excitations of gluon fields.
However, it has been a challenge for lattice QCD to quantify the
effect. In this work, we simulate QCD at high temperatures with chiral fermions.
The exact chiral symmetry enables us to
separate the contribution from the axial $U(1)$ breaking from others
among the susceptibilities in the scalar and pseudoscalar channels.
Our result in two-flavor QCD indicates that the chiral susceptibility,
 which is conventionally used as a probe for $SU(2)_L\times SU(2)_R$
breaking, is actually dominated by the axial $U(1)$ breaking at
temperatures $T\ge 165$ MeV.\\\\
Report numbers: OU-HET-1112, KEK-CP-0384, YITP-21-152
}

\FullConference{%
 The 38th International Symposium on Lattice Field Theory, LATTICE2021
  26th-30th July, 2021
  Zoom/Gather@Massachusetts Institute of Technology
}

\begin{document}
\maketitle

\section{Introduction}

It is widely believed that in the early universe
around 10 $\mu$s after the Big-Bang,
at a temperature around 150 MeV,
there has been a phase transition
where the quarks and gluons in the plasma got confined.
It is also believed that the $SU(2)_L\times SU(2)_R$ chiral symmetry
was spontaneously broken at the same temperature.
The signal of the so-called chiral phase transition
should be detected by the expectation value of the quark condensate,
which is zero at temperatures higher than the critical value $T_c$,
while it becomes non-zero at lower temperatures.
Simulating lattice QCD, the temperature and quark mass dependences
of the chiral condensate or its susceptibility have been intensively studied
\cite{Karsch:1994hm, Aoki:2006we, Cheng:2006qk, Bazavov:2011nk,
  Bhattacharya:2014ara, Bonati:2015bha, Brandt:2016daq, Taniguchi:2016ofw, Ding:2019prx}.

In this work, we focus on a fact that
the chiral condensate breaks not only $SU(2)_L\times SU(2)_R$
but also the axial $U(1)$ symmetry (or $U(1)_A$ in short).
Since the $U(1)_A$ symmetry is explicitly broken by anomaly,
which is true in any energy scale or temperature,
one may consider that the temperature and quark mass dependences
of the $U(1)_A$ anomaly should be milder than that of $SU(2)_L\times SU(2)_R$
and there should be little effect on the phase transition from it.

In the early days of QCD, however, they tried to explain
the spontaneous chiral symmetry breaking by the enhancement
of the $U(1)_A$ anomaly through the instanton effects.
For example, in Ref.~\cite{Callan:1977gz}, Callan, Dashen and Gross proposed
a scenario in which the chiral condensate is triggered by
the four fermion interaction induced by instantons (see also \cite{Diakonov:1984vw}).
If this scenario as well as its inverse are true,
the $SU(2)_L\times SU(2)_R$ symmetry is
recovered at the critical temperature $T_c$ as a consequence of
disappearance of the $U(1)_A$ anomaly and instantons.
It should also be noted that the disappearance of the $U(1)_A$ anomaly
would affect the universal property of the chiral phase transiton \cite{Pisarski:1983ms}.

It has been a challenge to examine the above two scenarios in QCD.
On the analytic side, the semi-classical treatment of
the QCD instantons is not good enough to describe the
low energy dynamics of QCD.
On the numerical side, in the conventional fermion formulation
in lattice QCD,
both of the $SU(2)_L\times SU(2)_R$ and $U(1)_A$ are
explicitly broken, and therefore, it is difficult to
disentangle the signals of their breaking/restoration
and their lattice artifacts.
Moreover, in our previous studies
\cite{Cossu:2015kfa,Tomiya:2016jwr, Aoki:2020noz}, 
we found that in the physical quantities related
to the $U(1)_A$  the lattice artifact is enhanced at
high temperatures, which makes the issue hard
even with M\"obius domain-wall fermions.

In this work, we study the chiral susceptibility
in two-flavor QCD with overlap fermions \cite{Neuberger:1997fp}.
The exact chiral symmetry with the Ginsparg-Wilson relation,
enables us to separate the $U(1)_A$ (in particular topological) effect from
others in a theoretically clean way.
Our analysis shows that the
signal of chiral susceptibility is dominated by
the $U(1)_A$ breaking effect rather than that of
$SU(2)_L\times SU(2)_R$ at high temperatures.
This result supports the scenario proposed in \cite{Callan:1977gz, Diakonov:1984vw}.
The analysis at $T\ge 190$ MeV was already
reported in our paper \cite{Aoki:2021qws}.
In this report, we add preliminary data at $T= 165$ MeV,
which is near the pseudo critical temperature at the physical quark masses.

\section{Chiral susceptibility and the $U(1)_A$ breaking contribution}

The $N_f$-flavor QCD partition function of the quark mass $m$ and temperature $T$
is given by a path integral
\begin{align}
  Z(m,T) = \int [dA] \det(D(A)+m)^{N_f}e^{-S_G(A,T)},
\end{align}
over the gluon field $A$, where the Dirac operator on quarks is denoted by $D(A)$
and the gluon action at temperature $T$ is by $S_G(A,T)$.
The chiral condensate and susceptibilty are given by its first and second
derivatives,
\begin{align}
  \Sigma(m,T)
  = \frac{1}{N_fV}\frac{\partial}{\partial m}\ln Z(m,T)\;\;\;\mbox{and}\;\;\;
  \chi(m,T)
  =  \frac{1}{N_fV}\frac{\partial^2}{\partial m^2}\ln Z(m,T),
\end{align}
 respectively.
 Here and in the following, we set $N_f=2$ and neglect the strange and heavier quarks.

\begin{figure*}[bth]
  \centering
    \includegraphics[width=14cm]{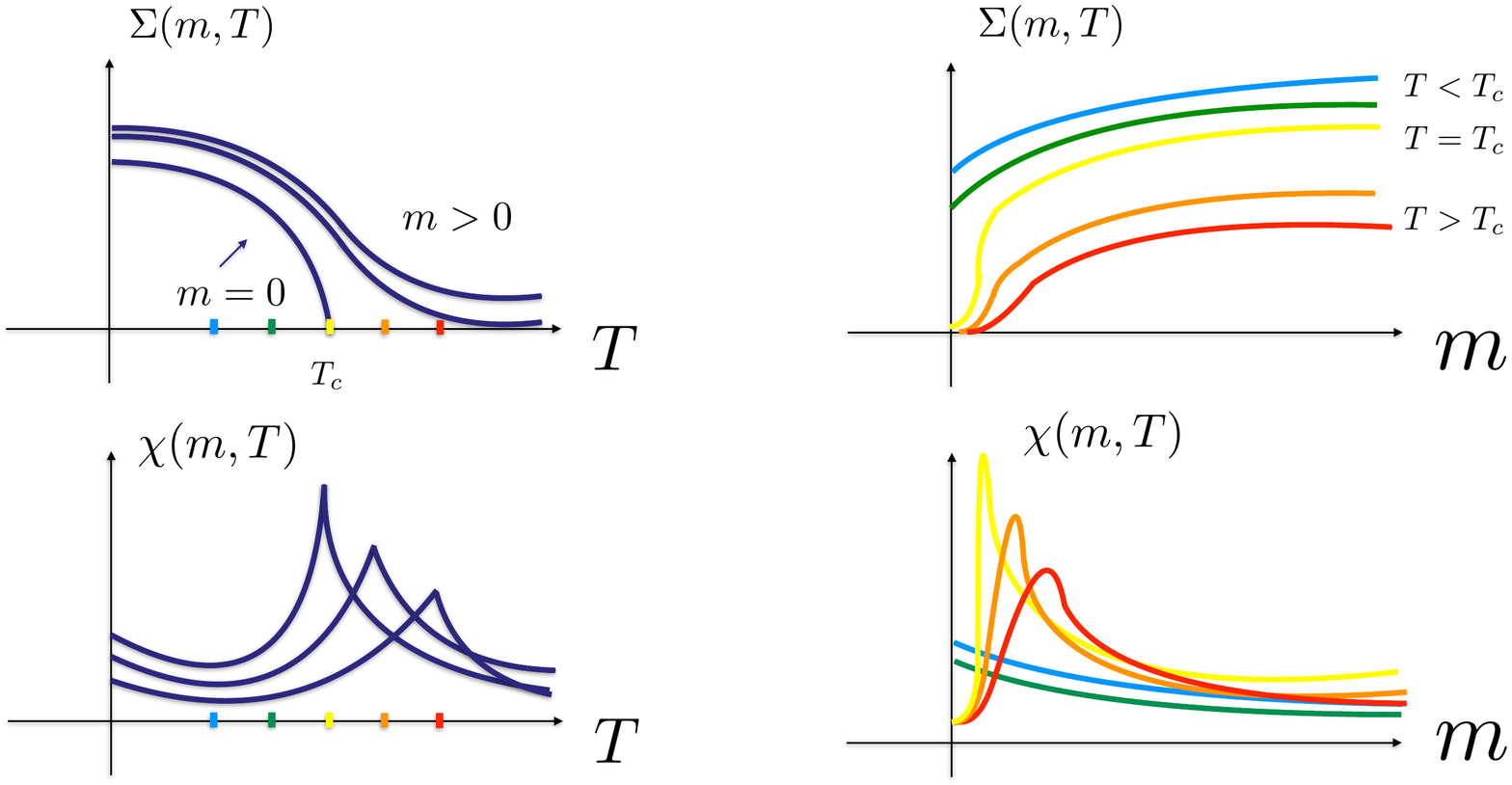}
  \caption{
    Schematic picture of the 
    chiral condensate (top panels) and susceitpibility (bottom).
    The temperature $T$ dependence (left panels) and 
    and that on the quark mass $m$ (right) at fixed temperatures
    marked by colored symbols are shown.
  }
  \label{fig:Tmdep}
\end{figure*}

 A schematic picture for their $T$ and $m$ dependences are
 shown in Fig.~\ref{fig:Tmdep}.
As presented in the top-left panel,
 the chiral condensate at $m=0$ vanishes at a critical temperature $T_c$,
 while the $T$ dependence gives a  milder crossover for the massive quarks\footnote{
   In Fig.~\ref{fig:Tmdep} the second or higher order phase transition is assumed.
   Even when the transition is the first order,
   the finite volume effect on the lattice would smooth the transition
   and the qualitative behavior of $\chi(m,T)$ would not change very much.
 }.
 Then the  chiral extrapolation at fixed temperatures should behave
 like the colored curves in the top-right panel.
 For $T<T_c$ the condensate converges to nonzero values, while
 it shows a sharp drop near the chiral limit when $T>T_c$.
 The drops of the chiral condensate should be reflected
 as peaks of the chiral susceptibility, as shown in the bottom-right panel.
 This peak is the sharpest and highest at $T=T_c$ and becomes 
 broader and lower, gradually moving to the right as $m$ increases.

 Since the fermion determinant is formally given by
 a product of eigemvalues of the Dirac operator,
 $\det(D(A)+m)=\prod_\lambda (i\lambda(A)+m)$,
 the chiral condensate can be expressed by
 a summation over the eigenvalues,
\begin{eqnarray}
\Sigma(m,T)
  &=\frac{1}{V}\left\langle \sum_\lambda \frac{1}{i\lambda(A)+m}\right\rangle.
\end{eqnarray}
The chiral susceptibility $\chi(m,T)=\chi^{\rm con.}(m,T) +\chi^{\rm dis.}(m,T)$ consists of the connected part $\chi^{\rm con.}(m,T)$,
which is a derivative of the chiral condensate with respect
to the valence quark mass,
and the disconnected part $\chi^{\rm dis.}(m,T)$, which is with respect to the sea quark mass.
These observables are given by the eigenvalues
only, and insensitive to the profile of the eigenfunctions.

From the Ward-Takahashi identity (WTI) of the $SU(2)_L\times SU(2)_R$ symmetry
and the anomalous WTI of the $U(1)_A$,
we can relate the chiral susceptibility
to scalar and pseudoscalar susceptibilities,
$\sum_x \langle S^i(x)S^i(0)\rangle$ and $\sum_x \langle P^i(x)P^i(0)\rangle$,
where the superscript $i=a$ for the triplet and $i=0$ for the singlet,
as well as the topological susceptibility $\chi_{\rm top.}(m,T)$ \cite{Aoki:2012yj, Nicola:2018vug, Nicola:2020smo}.
We can also show that only the connected part
contains a constant quadratic UV divergence,
which is the same as the one contained in $\Sigma(m,T)/m$.
Therefore, we define the subtracted condensate by
\begin{align}
  \frac{\Sigma_{\rm sub.}(m,T)}{m}\equiv \left[\frac{\Sigma(m,T)}{m}-\frac{\langle |Q(A)|\rangle}{m^2V}\right]
  -\left[\frac{\Sigma(m_{\rm ref},T)}{m_{\rm ref}}-\frac{\langle |Q(A)|\rangle|_{m=m_{\rm ref}}}{m^2_{\rm ref}V}\right],
  \end{align}
with a reference mass $m_{\rm ref}$.
Here possible chiral zero mode's effect is also subtracted
by introducing
the term with the topological charge $Q(A)$, which is
irrelevant to the UV divergence.

Specifically the subtracted connected chiral susceptibility is decomposed as
\begin{align}
\chi_{\rm sub.}^{\rm con.}(m,T)=\textcolor{black}{\underbrace{-\Delta_{U(1)}(m,T)+\frac{\langle |Q(A)|\rangle}{m^2V}}_{\mbox{$U(1)_A$ breaking}}
+\color{black}{\underbrace{\frac{\Sigma_{\rm sub.}(m,T)}{m}}_{\mbox{mixed}}}},
\end{align}
and the disconnected susceptibility is
\begin{align}
\chi^{\rm dis.}(m,T) =\textcolor{black}{\underbrace{\frac{N_f}{m^2}\chi_{\rm top.}(m,T)}_{\mbox{$U(1)_A$ breaking}}} 
+\textcolor{black}{\underbrace{\Delta_{SU(2)}^{(1)}(m,T)-\Delta_{SU(2)}^{(2)}(m,T)}_{\mbox{$SU(2)_L\times SU(2)_R$ breaking}}},
\end{align}
where $\Delta_{U(1)}(m,T)$ is the so-called $U(1)_A$ susceptibility:
\begin{align}
\Delta_{U(1)}(m,T) \equiv  
  \sum_x \langle P^a(x)P^a(0)-S^a(x)S^a(0)\rangle,
\end{align}
and $\Delta_{SU(2)}^{(1)}(m,T)$ and $\Delta_{SU(2)}^{(2)}(m,T)$ measure
$SU(2)_L\times SU(2)_R$ breakings,
\begin{align}
\Delta_{SU(2)}^{(1)}(m,T) &\equiv  
\sum_x \langle S^0(x)S^0(0)-P^a(x)P^a(0)\rangle,\\
\Delta_{SU(2)}^{(2)}(m,T) &\equiv  
  \sum_x \langle S^a(x)S^a(0)-P^0(x)P^0(0)\rangle.
\end{align}
Now our goal is to quantify how much the $U(1)_A$ breaking contributions
\begin{align}
  \chi_{A}^{\rm con.}(m,T) = -\Delta_{U(1)}(m,T)+\frac{\langle |Q(A)|\rangle}{m^2V} \;\;\;\mbox{and}\;\;\;
  \chi_A^{\rm dis.}(m,T) = \frac{N_f}{m^2}\chi_{\rm top.}(m,T)
\end{align}
dominate the signals.

So far we have used the continuum notation for the observables.
The lattice formulas are given using the eigenvalue $\lambda_m$ of
the massive overlap Dirac operator $H_m = \gamma_5 [(1-m)D_{ov}+m]$:
\begin{align}
\Delta_{U(1)}(m,T) &= \frac{1}{V(1-m^2)^2}\left\langle \sum_{\lambda_m}\frac{2m^2(1-\lambda_m^2)^2}{\lambda_m^4}\right\rangle,
  \\
  \Sigma(m,T)&=   \frac{1}{V(1-m^2)}\left\langle \sum_{\lambda_m} \frac{m(1-\lambda_m^2)}{\lambda_m^2}\right\rangle,\\
\chi^{\rm dis.}(m,T) &= 
  \frac{N_f}{V}\left[\frac{1}{(1-m^2)^2}
    \left\langle \left(\sum_{ \lambda_m} \frac{m(1-\lambda_m^2)}{\lambda_m^2}\right)^2\right\rangle
    -|\Sigma(m,T)|^2 V^2
    \right].
\end{align}
We emphasize that the exact chiral symmetry through the Ginsparg-Wilson relation
is essential in this spectral decomposition and the theoretically
clean and unambiguous separation of the $U(1)_A$ breaking effect.

\section{Lattice simulations}

We simulate $N_f=2$ lattice QCD with a lattice spacing $a=0.075$~fm.
We employ the Symanzik gauge action and M\"obius domain-wall fermion \cite{Brower:2005qw}
with the residual mass $m_{\rm res}\sim 1$ MeV for the configuration generations.
Since the physical observables related to the $U(1)_A$ anomaly at high temperatures
are sensitive to the lattice artifacts, we reweight the
configuration by the overlap fermion determinant.
The quark mass covers 3--30 MeV, where the lowest value
is below the physical point $\sim 4$ MeV.
The simulated temperatures\footnote{
The results at $T\sim 165$ MeV are preliminary.
  } are 165, 195, 220, 260, and 330 MeV,
which is tuned by the temporal size of the lattice $L_t=$8--16.
At $T=220$ MeV, we simulate 4 different volume size lattices with $L=1.8$--3.6~fm
to estimate the finite size systematics.

In the spectral decomposition of the chiral susceptibility,
we need to truncate the summation over the eigenvalues.
In this work, we cutoff the summation at 30--40-th lowest modes,
which corresponds to 150--300 MeV.
For $T\le 260$~MeV, we find a good saturation and consistency with direct inversion of M\"obius domain-wall Dirac operator,
but at $T=$330 MeV, the convergence is marginal and we use the direct inversion.

\begin{figure*}[bth]
  \centering
  \includegraphics[width=7cm]{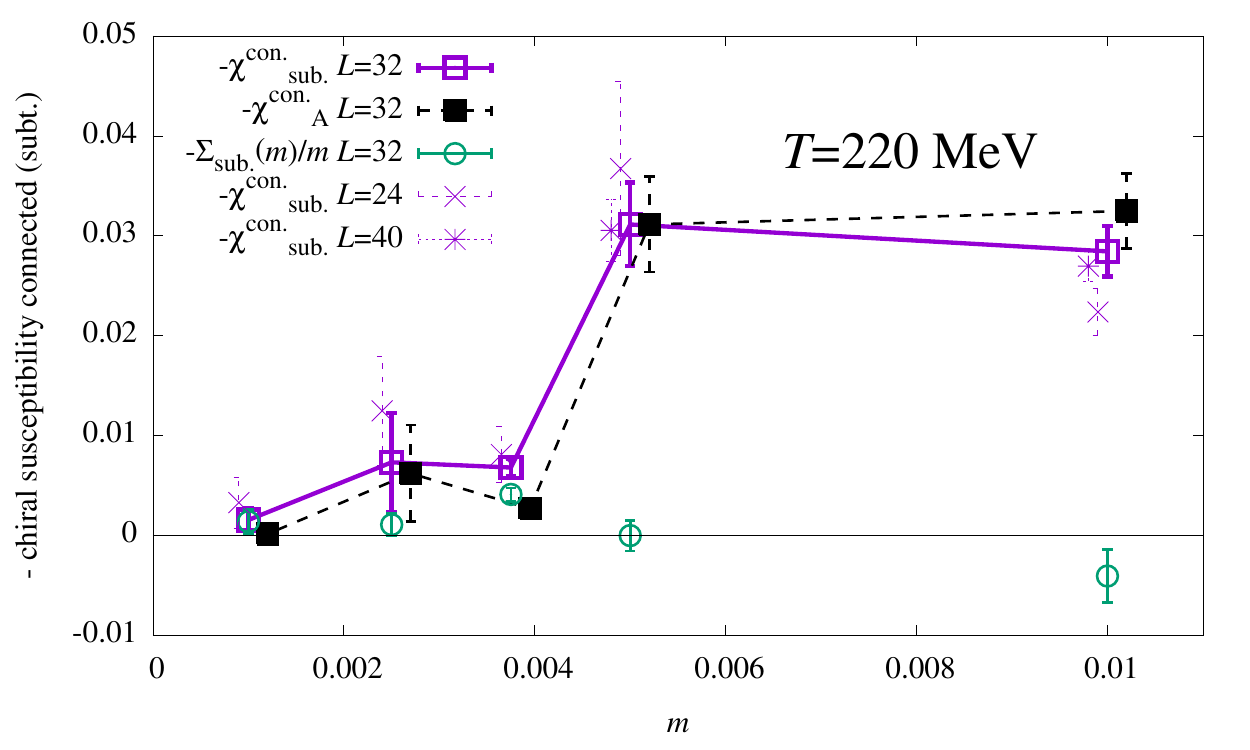}
  \includegraphics[width=7cm]{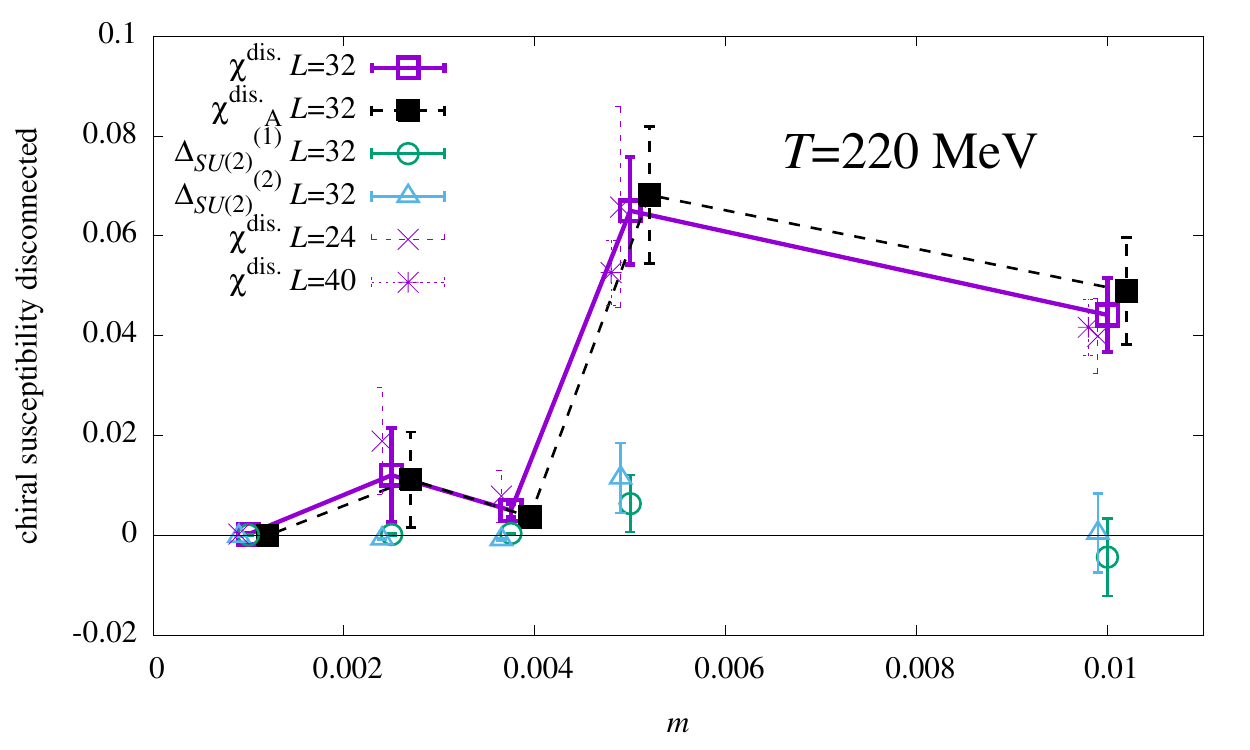}
  \caption{
    The connected (left panel) and disconnected (right) susceptibilities at $T=220$ MeV
    are plotted as a function of the quark mass.
    Compared to the total contribution in open mazenta squares,
    the axial $U(1)$ anomaly part in black filled squares are dominant.
    No sizable volume effect is seen.
  }
  \label{fig:result220}
  \centering
  \includegraphics[width=7cm]{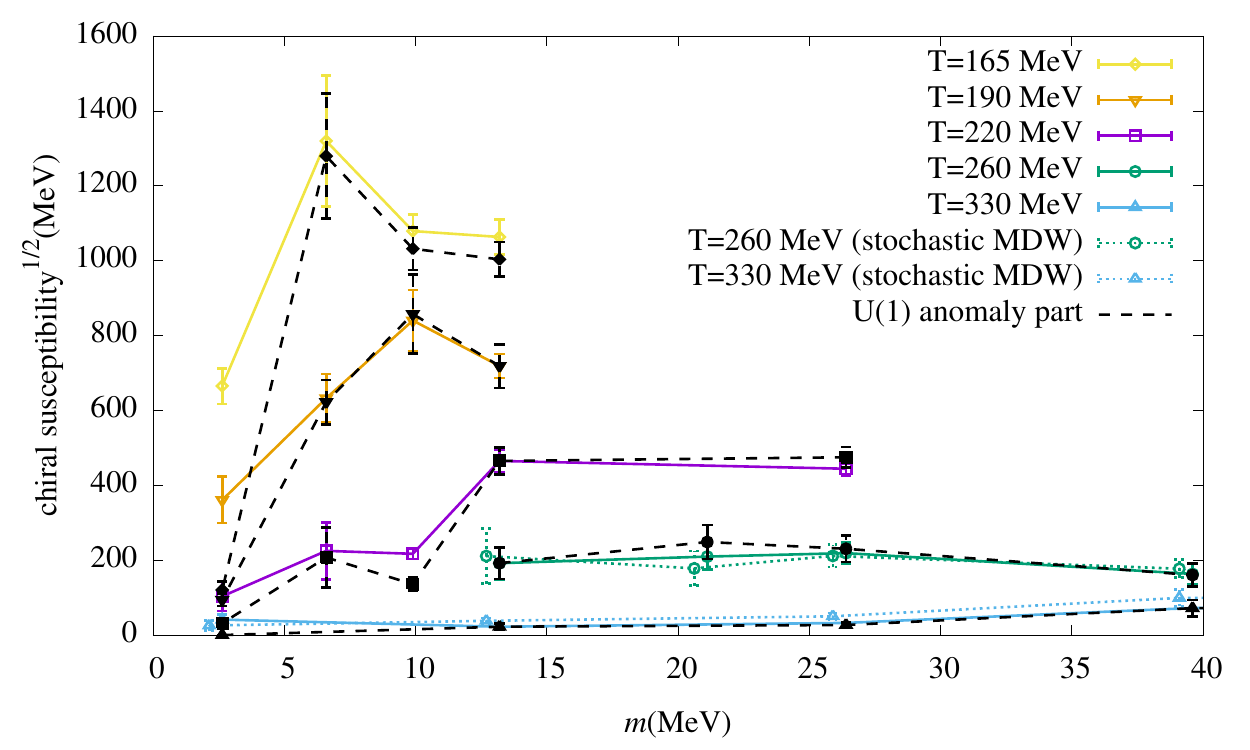}
  \includegraphics[width=7cm]{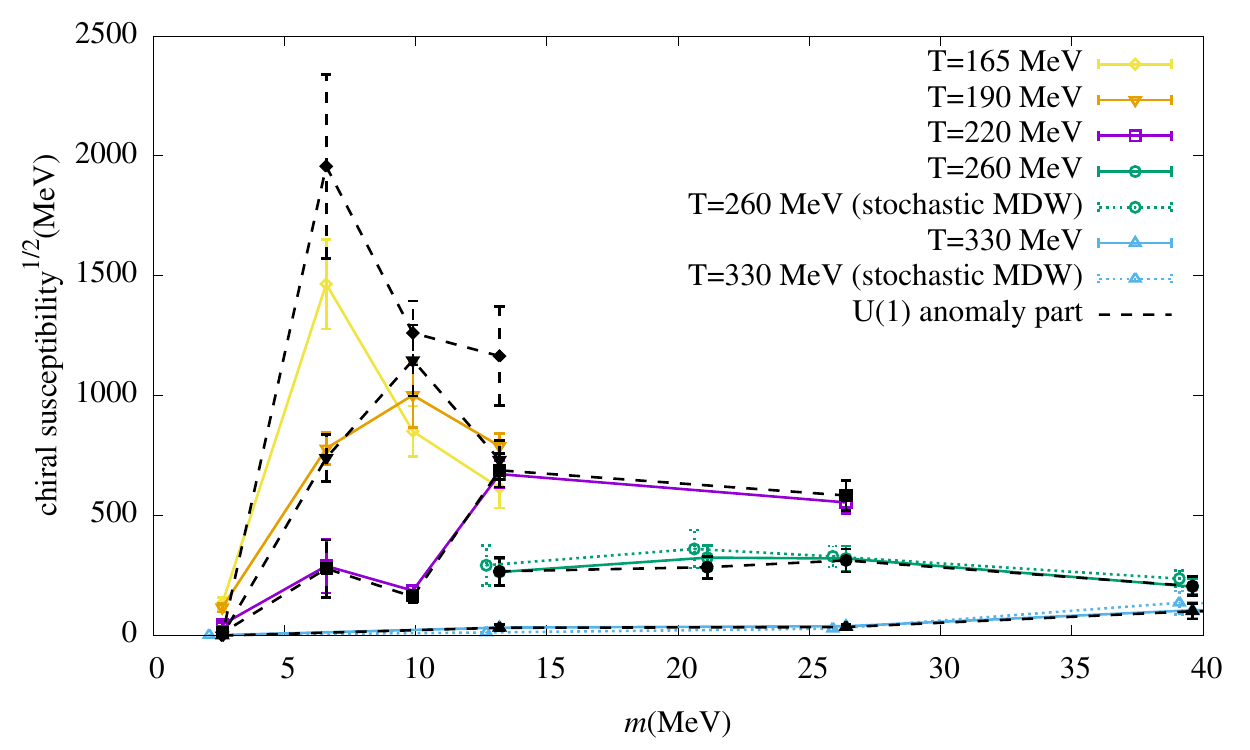}
  \caption{
    The connected (left panel) and disconnected (right) susceptibilities 
    are plotted as a function of the quark mass at 5 different temperatures.
    Compared to the total contribution in open colored symbols,
the axial $U(1)$ anomaly part in black filled are dominant.
  }
  \label{fig:result-summary}
\end{figure*}

In Fig.~\ref{fig:result220}, the (subtracted) connected susceptibility at $T=220$ MeV
is plotted as a function of the quark mass in the left panel
and that for the disconnected part at the same temperature is shown in the right panel.
Compared to the total contribution in open mazenta squares,
the axial $U(1)$ anomaly part in black filled squares are dominant.
To be specific, the connected part is dominated by axial $U(1)$ susceptibility 
and disconnected one is governed by the topological susceptibiliy.
The data are obtained on 3 different volume lattices and they are all consistent.

The dominance by the $U(1)$ breaking is seen at 5 simulated temperatures above $T_c$,
which is presented in Fig.~\ref{fig:result-summary}.
Here the data at 165 MeV are still preliminary,
while others are given in Ref.~\cite{Aoki:2021qws}.
Here the colored open symbols are our data for the chiral susceptibility
while the black filled ones are the axial $U(1)$ breaking part.
In fact, $\sim 90$\% of signals comes from the axial $U(1)$ breaking.
It is interesting to note that the behavior of the peaks, which is highest at $T=165$ MeV
and getting lower, shifting to the higher quark mass point with $T$,
is consistent with a naive picture presented in the introduction.
We also remark that the chiral limit of the axial $U(1)$ breaking is strongly suppressed.

\section{Summary}

The chiral condensate and susceptibility are related to both of $SU(2)_L\times SU(2)_R$ and $U(1)_A$.
In the spectral decomposition of the Dirac operator with exact chiral symmetry on the lattice,
we can separate the purely $U(1)_A$ breaking effect.
We have found that the chiral susceptibility is dominated by the $U(1)_A$ breaking contribution
at $T\gtrsim T_c$. Specifically, the connected part is described by the  axial $U(1)$ susceptibility,
while the disconnected part is dominated by the  topological susceptibility (divided by $m^2$).
Our result indicates that the chiral susceptibility is probing the $m$ and $T$ dependences
of the $U(1)_A$ breaking, rather than that of $SU(2)_L\times SU(2)_R$.
The axial $U(1)$ anomaly may play more important role in the QCD phase diagram than expected.

We thank 
H.-T. Ding, C. Gattringer, L. Glozman, and S. Takeda
for useful discussions.
We thank P. Boyle for correspondence for starting simulation with Grid 
and I. Kanamori for helping us on the simulations on K computer with Bridge++.
We also thank the members of JLQCD collaboration for their encouragement and support.
We thank the Yukawa Institute for Theoretical Physics at Kyoto University.
Discussions during the YITP workshop YITP-W-20-08 on "Progress in Particle Physics 2020" were useful to complete this work.
Numerical simulations were performed using the QCD software packages Iroiro++ \cite{Cossu:2013ola},
   Grid \cite{Boyle:2015tjk}, and Bridge++ \cite{Ueda:2014rya}
on IBM System Blue Gene Solution at KEK under a
support of its Large Scale Simulation Program (No. 16/17-14) and Oakforest-PACS at JCAHPC
under a support of the HPCI System Research Projects (Project IDs: hp170061, hp180061,
hp190090, and hp200086), Multidisciplinary Cooperative Research Program in CCS, University of Tsukuba
(Project IDs: xg17i032 and xg18i023) and K computer provided by the RIKEN Center for Computational Science.
This work is supported in part by the Japanese Grant-in-Aid for Scientific Research
(No. JP26247043, JP16H03978, JP18H01216, JP18H03710, JP18H04484, JP18H05236), and by MEXT as
“Priority Issue on Post-K computer" (Elucidation of the Fundamental Laws and Evolution of the
Universe) and by Joint Institute for Computational Fundamental Science (JICFuS).


\begin{thebibliography}{99}
  
\bibitem{Karsch:1994hm}
F.~Karsch and E.~Laermann,
``Susceptibilities, the specific heat and a cumulant in two flavor QCD,''
Phys. Rev. D \textbf{50}, 6954-6962 (1994)
doi:10.1103/PhysRevD.50.6954
[arXiv:hep-lat/9406008 [hep-lat]].

\bibitem{Aoki:2006we}
Y.~Aoki, G.~Endrodi, Z.~Fodor, S.~D.~Katz and K.~K.~Szabo,
``The Order of the quantum chromodynamics transition predicted by the standard model of particle physics,''
Nature \textbf{443}, 675-678 (2006)
doi:10.1038/nature05120
[arXiv:hep-lat/0611014 [hep-lat]].

\bibitem{Cheng:2006qk}
  M.~Cheng  \textit{et al.},
``The Transition temperature in QCD,''
Phys. Rev. D \textbf{74}, 054507 (2006)
doi:10.1103/PhysRevD.74.054507
[arXiv:hep-lat/0608013 [hep-lat]].

\bibitem{Bazavov:2011nk}
  A.~Bazavov \textit{et al.} [HotQCD Collaboration],
``The chiral and deconfinement aspects of the QCD transition,''
Phys. Rev. D \textbf{85}, 054503 (2012)
doi:10.1103/PhysRevD.85.054503
[arXiv:1111.1710 [hep-lat]].


\bibitem{Bhattacharya:2014ara}
  T.~Bhattacharya \textit{et al.} [HotQCD Collaboration],
``QCD Phase Transition with Chiral Quarks and Physical Quark Masses,''
Phys. Rev. Lett. \textbf{113}, no.8, 082001 (2014)
doi:10.1103/PhysRevLett.113.082001
[arXiv:1402.5175 [hep-lat]].


\bibitem{Bonati:2015bha}
C.~Bonati, M.~D'Elia, M.~Mariti, M.~Mesiti, F.~Negro and F.~Sanfilippo,
``Curvature of the chiral pseudocritical line in QCD: Continuum extrapolated results,''
Phys. Rev. D \textbf{92}, no.5, 054503 (2015)
doi:10.1103/PhysRevD.92.054503.


\bibitem{Brandt:2016daq}
B.~B.~Brandt, A.~Francis, H.~B.~Meyer, O.~Philipsen, D.~Robaina and H.~Wittig,
``On the strength of the $U_A(1)$ anomaly at the chiral phase transition in $N_f=2$ QCD,''
JHEP \textbf{12}, 158 (2016)
doi:10.1007/JHEP12(2016)158
[arXiv:1608.06882 [hep-lat]].


\bibitem{Taniguchi:2016ofw}
  Y.~Taniguchi \textit{et al.} [WHOT-QCD Collaboration],
``Exploring $N_{f}$ = 2+1 QCD thermodynamics from the gradient flow,''
Phys. Rev. D \textbf{96}, no.1, 014509 (2017)
[erratum: Phys. Rev. D \textbf{99}, no.5, 059904 (2019)]
doi:10.1103/PhysRevD.96.014509
[arXiv:1609.01417 [hep-lat]].

\bibitem{Ding:2019prx}
  H.~T.~Ding \textit{et al.} [HotQCD Collaboration],
``Chiral Phase Transition Temperature in ( 2+1 )-Flavor QCD,''
Phys. Rev. Lett. \textbf{123}, no.6, 062002 (2019)
doi:10.1103/PhysRevLett.123.062002
[arXiv:1903.04801 [hep-lat]].



\bibitem{Callan:1977gz}
C.~G.~Callan, Jr., R.~F.~Dashen and D.~J.~Gross,
``Toward a Theory of the Strong Interactions,''
Phys. Rev. D \textbf{17}, 2717 (1978)
doi:10.1103/PhysRevD.17.2717

\bibitem{Diakonov:1984vw}
D.~Diakonov and V.~Y.~Petrov,
``CHIRAL CONDENSATE IN THE INSTANTON VACUUM,''
Phys. Lett. B \textbf{147}, 351-356 (1984)
doi:10.1016/0370-2693(84)90132-1



\bibitem{Pisarski:1983ms} 
  R.~D.~Pisarski and F.~Wilczek,
  ``Remarks on the Chiral Phase Transition in Chromodynamics,''
  Phys.\ Rev.\ D {\bf 29}, 338 (1984)
  doi:10.1103/PhysRevD.29.338.
 


  
\bibitem{Cossu:2015kfa} 
  G.~Cossu {\it et al.} [JLQCD Collaboration],
  ``Violation of chirality of the M\"obius domain-wall Dirac operator from the eigenmodes,''
  Phys.\ Rev.\ D {\bf 93}, no. 3, 034507 (2016)
  doi:10.1103/PhysRevD.93.034507
  [arXiv:1510.07395 [hep-lat]].


  
\bibitem{Tomiya:2016jwr} 
  A.~Tomiya {\it et al.} [JLQCD Collaboration]
  ``Evidence of effective axial U(1) symmetry restoration at high temperature QCD,''
  Phys.\ Rev.\ D {\bf 96}, no. 3, 034509 (2017)
  Addendum: [Phys.\ Rev.\ D {\bf 96}, no. 7, 079902 (2017)]
  doi:10.1103/PhysRevD.96.034509, 10.1103/PhysRevD.96.079902
  [arXiv:1612.01908 [hep-lat]].


\bibitem{Aoki:2020noz}
S.~Aoki \textit{et al.} [JLQCD],
``Study of the axial $U(1)$ anomaly at high temperature with lattice chiral fermions,''
Phys. Rev. D \textbf{103}, no.7, 074506 (2021)
doi:10.1103/PhysRevD.103.074506
[arXiv:2011.01499 [hep-lat]].
  


\bibitem{Neuberger:1997fp} 
  H.~Neuberger,
  ``Exactly massless quarks on the lattice,''
  Phys.\ Lett.\ B {\bf 417}, 141 (1998)
  doi:10.1016/S0370-2693(97)01368-3
  [hep-lat/9707022].
  


\bibitem{Aoki:2021qws}
S.~Aoki \textit{et al.} [JLQCD collaboration],
``Role of axial U(1) anomaly in chiral susceptibility of QCD at high temperature,''
[arXiv:2103.05954 [hep-lat]].


  




\bibitem{Aoki:2012yj} 
  S.~Aoki, H.~Fukaya and Y.~Taniguchi,
  ``Chiral symmetry restoration, eigenvalue density of Dirac operator and axial U(1) anomaly at finite temperature,''
  Phys.\ Rev.\ D {\bf 86}, 114512 (2012)
  doi:10.1103/PhysRevD.86.114512
  [arXiv:1209.2061 [hep-lat]].


\bibitem{Nicola:2018vug}
A.~G\'omez Nicola and J.~Ruiz De Elvira,
``Chiral and $U(1)_A$ restoration for the scalar and pseudoscalar meson nonets,''
Phys. Rev. D \textbf{98}, no.1, 014020 (2018)
doi:10.1103/PhysRevD.98.014020
[arXiv:1803.08517 [hep-ph]].
  

\bibitem{Nicola:2020smo}
A.~G.~Nicola,
``Light quarks at finite temperature: chiral restoration and the fate of the $U(1)_A$ symmetry,''
Eur. Phys. J. ST \textbf{230}, no.6, 1645-1657 (2021)
doi:10.1140/epjs/s11734-021-00147-4
[arXiv:2012.13809 [hep-ph]].

\bibitem{Brower:2005qw} 
  R.~C.~Brower, H.~Neff and K.~Orginos,
  ``Mobius fermions,''
  Nucl.\ Phys.\ Proc.\ Suppl.\  {\bf 153}, 191 (2006)
  doi:10.1016/j.nuclphysbps.2006.01.047
  [hep-lat/0511031];
  ``The Möbius domain wall fermion algorithm,''
  Comput.\ Phys.\ Commun.\  {\bf 220}, 1 (2017)
  doi:10.1016/j.cpc.2017.01.024
  [arXiv:1206.5214 [hep-lat]].


\bibitem{Cossu:2013ola}
G.~Cossu, J.~Noaki, S.~Hashimoto, T.~Kaneko, H.~Fukaya, P.~A.~Boyle and J.~Doi,
``JLQCD IroIro++ lattice code on BG/Q,''
doi:https://doi.org/10.22323/1.187.0482
[arXiv:1311.0084 [hep-lat]].

\bibitem{Boyle:2015tjk}
P.~Boyle, A.~Yamaguchi, G.~Cossu and A.~Portelli,
``Grid: A next generation data parallel C++ QCD library,''
doi:https://doi.org/10.22323/1.251.0023
[arXiv:1512.03487 [hep-lat]].


\bibitem{Ueda:2014rya}
S.~Ueda, S.~Aoki, T.~Aoyama, K.~Kanaya, H.~Matsufuru, S.~Motoki, Y.~Namekawa, H.~Nemura, Y.~Taniguchi and N.~Ukita,
``Development of an object oriented lattice QCD code 'Bridge++',''
J. Phys. Conf. Ser. \textbf{523}, 012046 (2014)
doi:10.1088/1742-6596/523/1/012046

\end{thebibliography}
\end{document}